\documentclass[12pt]{article}
\usepackage{amsmath,amssymb,amsthm,amsxtra,overpic,bbm,bm,epsfig,ulem}
\usepackage{color}
\usepackage{cite}

\usepackage{hyperref}
\usepackage{url}

\def\thefootnote{\fnsymbol{footnote}}

\begin{document}

\vspace{0.2cm}

\begin{center}
{\large\bf Bridging resonant leptogenesis and low-energy CP violation \\
with an RGE-modified seesaw relation}
\end{center}

\vspace{0.2cm}

\begin{center}
{\bf Zhi-zhong Xing$^{1,2}$}
\footnote{E-mail: xingzz@ihep.ac.cn}
and {\bf Di Zhang$^{1}$}
\footnote{E-mail: zhangdi@ihep.ac.cn (corresponding author)}
\\
{\small $^{1}$Institute of High Energy Physics and School of Physical Sciences, \\
University of Chinese Academy of Sciences, Beijing 100049, China \\
$^{2}$Center of High Energy Physics, Peking University, Beijing 100871, China}
\end{center}

\vspace{2cm}
\begin{abstract}
We propose a special type-I seesaw scenario in which the Yukawa coupling matrix
$Y^{}_\nu$ can be fully reconstructed by using the light Majorana neutrino
masses $m^{}_i$, the heavy Majorana neutrino masses $M^{}_i$ and the PMNS lepton
flavor mixing matrix $U$. It is the RGE-induced correction to the seesaw relation
that helps interpret the observed baryon-antibaryon asymmetry of the Universe
via flavored resonant thermal leptogenesis with $M^{}_1 \simeq M^{}_2 \ll M^{}_3$.
We show that our idea works well in either the $\tau$-flavored regime with
equilibrium temperature $T \simeq M^{}_1 \in (10^9, 10^{12}]$ GeV or
the $(\mu+\tau)$-flavored regime with $T \simeq M^{}_1 \in (10^5, 10^9]$ GeV,
provided the light neutrinos have a normal mass ordering.
We find that the same idea is also viable for a {\it minimal} type-I seesaw model
with two nearly degenerate heavy Majorana neutrinos.
\end{abstract}

\newpage

\def\thefootnote{\arabic{footnote}}
\setcounter{footnote}{0}
\setcounter{figure}{0}

\section{Introduction}

A special bonus of the canonical (type-I) seesaw mechanism
\cite{Minkowski:1977sc,Yanagida:1979as,GellMann:1980vs,Glashow:1979nm,Mohapatra:1979ia}
is the thermal leptogenesis mechanism \cite{Fukugita:1986hr}, which provides an elegant
way to interpret the mysterious matter-antimatter asymmetry of our Universe. The
key points of these two correlated mechanisms can be summed up in one sentence:
the tiny masses of three known neutrinos $\nu^{}_i$ are ascribed to the existence of
three heavy Majorana neutrinos $N^{}_i$ (for $i = 1, 2, 3$), whose lepton-number-violating
and CP-violating decays result in a net lepton-antilepton number asymmetry
$Y^{}_{\rm L}$ which is finally converted to a net baryon-antibaryon number asymmetry
$Y^{}_{\rm B}$ as observed today.

In the standard model (SM) extended with three right-handed neutrinos
and lepton number violation, it is the following {\it seesaw} formula that bridges the gap
between the masses of $\nu^{}_i$ (denoted as $m^{}_i$) and those of $N^{}_i$ (denoted
as $M^{}_i$):
\begin{eqnarray}
M^{}_\nu = -v^2 \left(Y^{}_\nu M^{-1}_{\rm R} Y^T_\nu\right) \; ,
%     (1)
\end{eqnarray}
where $M^{}_\nu$ represents the light (left-handed) Majorana neutrino mass matrix,
$v \simeq 174$ GeV is the vacuum expectation value of the SM neutral Higgs field,
$M^{}_{\rm R}$ stands for the heavy (right-handed) Majorana neutrino mass matrix,
and $Y^{}_\nu$ is a dimensionless coupling matrix describing the strength of Yukawa
interactions between the Higgs and neutrino fields. The eigenvalues of $M^{}_\nu$
(i.e., $m^{}_i$) can be strongly suppressed by those of $M^{}_{\rm R}$ (i.e.,
$M^{}_i$) as a consequence of $M^{}_i \gg v$ (for $i=1,2,3$), and that is
why $m^{}_i \ll v$ naturally holds.

Although such a seesaw picture is qualitatively attractive, it cannot make any
quantitative predictions unless the textures of $M^{}_{\rm R}$ and $Y^{}_\nu$ are
fully determined \cite{Xing:2019vks}. Without loss of generality, one may always
take the basis in which both the charged-lepton mass matrix $M^{}_l$ and the heavy
Majorana neutrino mass matrix $M^{}_{\rm R}$ are diagonal (i.e.,
$M^{}_l = D^{}_l \equiv {\rm Diag}\{m^{}_e, m^{}_\mu, m^{}_\tau\}$ and
$M^{}_{\rm R} = D^{}_N \equiv \{M^{}_1, M^{}_{2}, M^{}_3\}$). In this case the
undetermined Yukawa coupling matrix $Y^{}_\nu$ can be parametrized as follows
--- the so-called Casas-Ibarra (CI) parametrization \cite{Casas:2001sr}:
\begin{eqnarray}
Y^{}_\nu = \frac{\rm i}{v} \hspace{0.05cm} U \sqrt{D^{}_\nu} \hspace{0.1cm} O
\sqrt{D^{}_N} \; ,
%     (2)
\end{eqnarray}
where $U$ is the Pontecorvo-Maki-Nakagawa-Sakata (PMNS) neutrino
mixing matrix \cite{Pontecorvo:1957cp,Maki:1962mu,Pontecorvo:1967fh}
used to diagonalize $M^{}_\nu$ in the chosen basis (i.e.,
$U^\dagger M^{}_\nu U^* = D^{}_\nu \equiv {\rm Diag}\{m^{}_1, m^{}_2, m^{}_3\}$),
and $O$ is an arbitrary complex orthogonal matrix. This popular parametrization
of $Y^{}_\nu$ is fully compatible with the seesaw formula in Eq.~(1), but the
arbitrariness of $O$ remains unsolved.

Note that it is the complex phases hidden in $Y^{}_\nu$ that govern the CP-violating
asymmetries $\varepsilon^{}_{i\alpha}$ between the lepton-number-violating
decays $N^{}_i \to \ell^{}_\alpha + H$ and $N^{}_i \to \overline{\ell^{}_\alpha}
+ \overline{H}$ (for $i=1,2,3$ and $\alpha = e, \mu, \tau$) \cite{Fukugita:1986hr,Luty:1992un,Covi:1996wh,Plumacher:1996kc}.
In particular, the {\it flavored} asymmetries
$\varepsilon^{}_{i\alpha}$ depend on both
$\left(Y^{*}_\nu\right)^{}_{\alpha i} \left(Y^{}_\nu\right)^{}_{\alpha j}$
and $\left(Y^{\dagger}_\nu Y^{}_\nu\right)^{}_{ij}$ (for $j\neq i = 1,2,3$),
but the {\it unflavored} asymmetries
$\varepsilon^{}_{i} \equiv \varepsilon^{}_{i e} + \varepsilon^{}_{i \mu}
+ \varepsilon^{}_{i \tau}$ are only dependent upon
$\left(Y^{\dagger}_\nu Y^{}_\nu\right)^{}_{ij}$
\cite{Barbieri:1999ma,Endoh:2003mz,Giudice:2003jh,Nardi:2006fx,Abada:2006ea,
Blanchet:2006be,Abada:2006fw,Dev:2017trv}. Given the CI parametrization of $Y^{}_\nu$ in Eq.~(2), one
can immediately see that $\varepsilon^{}_{i}$ have nothing to do with the
PMNS matrix $U$ \cite{Xing:2009vb,Rodejohann:2009cq,Antusch:2009gn},
while $\varepsilon^{}_{i\alpha}$ will depend directly on $U$ if $O$ is assumed
to be real \cite{Pascoli:2003uh,Pascoli:2006ie,Pascoli:2006ci,Branco:2006ce,
Branco:2011zb,Moffat:2018smo}.

Note also that both $U$ and $D^{}_\nu$ in Eq.~(2) are defined at the
seesaw scale $\Lambda^{}_{\rm SS} \gg v$,
which can be related to their counterparts at the Fermi scale
$\Lambda^{}_{\rm EW} \sim v$ via the one-loop renormalization-group equations (RGEs)
\cite{Chankowski:1993tx,Babu:1993qv,Antusch:2001ck,Antusch:2005gp,Mei:2005qp,
Ohlsson:2013xva}. In this connection the RGE-induced correction to the CI
parametrization of $Y^{}_\nu$ has recently been taken into account \cite{Xing:2020erm}
%%%%%%%%%%%%%%%%%%%%%%%%%%%%%%%%%%%%%%%%%%%%%%%%%%%%%%%%%%%%%%%%%%%%
\footnote{A similar RGE-modified CI parametrization of $Y^{}_\nu$ has been
given in the case of the minimal supersymmetric standard model (MSSM)
extended with the seesaw mechanism \cite{Zhao:2020bzx}.}:
%%%%%%%%%%%%%%%%%%%%%%%%%%%%%%%%%%%%%%%%%%%%%%%%%%%%%%%%%%%%%%%%%%%%%%
\begin{eqnarray}
Y^{}_\nu \left( \Lambda^{}_{\rm SS} \right) = \frac{\rm i}{v} \hspace{0.05cm}
I^{}_0 \hspace{0.05cm} T^{}_l \hspace{0.05cm} U \left( \Lambda^{}_{\rm EW} \right)
\sqrt{D^{}_\nu \left( \Lambda^{}_{\rm EW}\right)} \hspace{0.1cm} O
\sqrt{D^{}_N \left( \Lambda^{}_{\rm SS} \right)} \; ,
%   (3)
\end{eqnarray}
where $T^{}_l = {\rm Diag}\{I^{}_e, I^{}_\mu, I^{}_\tau\}$, and the evolution
functions $I^{}_0$ and $I^{}_\alpha$ (for $\alpha = e, \mu, \tau$) are given by
\begin{eqnarray}
I^{}_0 \hspace{-0.2cm} & = & \hspace{-0.2cm}
\exp{\left[ -\frac{1}{32\pi^2} \int^{\ln{(\Lambda^{}_{\rm SS}
/\Lambda^{}_{\rm EW})}}_{0} \left[ 3 g^2_2(t) - 6 y^2_t(t) - \lambda(t)
\right] {\rm d}t \right]} \;,
\nonumber \\
I^{}_\alpha \hspace{-0.2cm} & = & \hspace{-0.2cm}
\exp{\left[ -\frac{3}{32\pi^2} \int^{\ln{
(\Lambda^{}_{\rm SS}/\Lambda^{}_{\rm EW})}}_{0} y^{2}_\alpha (t)
\hspace{0.05cm} {\rm d}t \right]} \;
%   (4)
\end{eqnarray}
in the SM with $g^{}_2$, $\lambda$, $y^{}_t$ and $y^{}_\alpha$ standing respectively
for the ${\rm SU(2)^{}_{L}}$ gauge coupling, the Higgs self-coupling constant,
the top-quark and charged-lepton Yukawa coupling eigenvalues \cite{Xing:2020erm}.
Eq.~(3) tells us that the unflavored CP-violating
asymmetries $\varepsilon^{}_{i}$ should also have something to do with the
PMNS matrix $U$ at low energies because of a slight departure of $T^{}_l$ from
the identity matrix. This new observation makes it possible to establish a
direct link between {\it unflavored} thermal leptogenesis and low-energy CP violation
under the assumption that $O$ is a real matrix \cite{Xing:2020erm,Zhao:2020bzx},
but one may still frown on the uncertainties associated with $O$.

In this work we simply assume the unconstrained orthogonal matrix $O$ to be the
identity matrix (i.e., $O = {\bf 1}$),
so as to reconstruct the Yukawa coupling matrix $Y^{}_\nu$ in
terms of not only $M^{}_i$ at the seesaw scale but also $m^{}_i$ and $U$ at low
energies. Considering the fact of $y^2_e \ll y^2_\mu \ll y^2_\tau \ll 1$ in the
SM, we find that $I^{}_e \simeq I^{}_\mu \simeq 1$ and
$I^{}_\tau \simeq 1 + \Delta^{}_\tau$ are two excellent approximations, where
\begin{eqnarray}
\Delta^{}_\tau = -\frac{3}{32\pi^2} \int^{\ln{(\Lambda^{}_{\rm SS}
/\Lambda^{}_{\rm EW})}}_{0} y^{2}_\tau (t) \hspace{0.05cm} {\rm d}t \;
%   (5)
\end{eqnarray}
denotes the small $\tau$-flavored effect \cite{Xing:2020erm}.
Then the expression of $Y^{}_\nu$ in Eq.~(3) can be somewhat simplified and
explicitly written as
\begin{eqnarray}
Y^{}_\nu \hspace{-0.2cm} & = & \hspace{-0.2cm}
\frac{\rm i}{v} \hspace{0.05cm} I^{}_0 \left[\left(\begin{matrix}
\sqrt{m^{}_1 M^{}_1} \hspace{0.05cm} U^{}_{e 1} &
\sqrt{m^{}_2 M^{}_2} \hspace{0.05cm} U^{}_{e 2} &
\sqrt{m^{}_3 M^{}_3} \hspace{0.05cm} U^{}_{e 3} \cr
\sqrt{m^{}_1 M^{}_1} \hspace{0.05cm} U^{}_{\mu 1} &
\sqrt{m^{}_2 M^{}_2} \hspace{0.05cm} U^{}_{\mu 2} &
\sqrt{m^{}_3 M^{}_3} \hspace{0.05cm} U^{}_{\mu 3} \cr
\sqrt{m^{}_1 M^{}_1} \hspace{0.05cm} U^{}_{\tau 1} &
\sqrt{m^{}_2 M^{}_2} \hspace{0.05cm} U^{}_{\tau 2} &
\sqrt{m^{}_3 M^{}_3} \hspace{0.05cm} U^{}_{\tau 3} \cr
\end{matrix}\right) \right.
\nonumber \\
\hspace{-0.2cm} & & \hspace{-0.15cm} + \left.
\Delta^{}_\tau \left(\begin{matrix}
0 & 0 & 0 \cr 0 & 0 & 0 \cr
\sqrt{m^{}_1 M^{}_1} \hspace{0.05cm} U^{}_{\tau 1} &
\sqrt{m^{}_2 M^{}_2} \hspace{0.05cm} U^{}_{\tau 2} &
\sqrt{m^{}_3 M^{}_3} \hspace{0.05cm} U^{}_{\tau 3} \cr
\end{matrix}\right) \right] \; ,
%   (6)
\end{eqnarray}
in which the scale indices $\Lambda^{}_{\rm SS}$ and $\Lambda^{}_{\rm EW}$
have been omitted for the sake of simplicity, but one should keep in mind
that the values of $m^{}_i$ and $U^{}_{\alpha i}$ (for $i=1,2,3$ and
$\alpha = e, \mu, \tau$) are subject to the Fermi scale $\Lambda^{}_{\rm EW}$.
With much less arbitrariness, we are going to show that such a special RGE-modified
seesaw scenario allows us to account for the observed baryon-to-photon ratio
$\eta \equiv n^{}_{\rm B}/n^{}_\gamma \simeq \left(6.12 \pm 0.03\right)
\times 10^{-10} \simeq 7.04 Y^{}_{\rm B}$ in today's Universe \cite{Aghanim:2018eyx}
by means of {\it flavored} {\it resonant} thermal leptogenesis with
$M^{}_1 \simeq M^{}_2 \ll M^{}_3$
\cite{Pilaftsis:1997dr,Pilaftsis:1997jf,Pilaftsis:2003gt,Anisimov:2005hr}
%%%%%%%%%%%%%%%%%%%%%%%%%%%%%%%%%%%%%%%%%%%%%%%%%%%%%%%%%%%%%%%%%%%%%%%%
\footnote{For such a heavy Majorana neutrino mass spectrum,
the role of $N^{}_3$ in thermal leptogenesis is expected to be negligible
because its contribution has essentially been washed out at $T \simeq M^{}_1
\simeq M^{}_2$.}.
%%%%%%%%%%%%%%%%%%%%%%%%%%%%%%%%%%%%%%%%%%%%%%%%%%%%%%%%%%%%%%%%%%%%%%%%
We find that our idea works in either the $\tau$-flavored regime with
equilibrium temperature $T \simeq M^{}_1 \in (10^9, 10^{12}]$ GeV or
the $(\mu+\tau)$-flavored regime with $T \simeq M^{}_1 \in (10^5, 10^9]$ GeV,
if the mass spectrum of three light Majorana neutrinos has a normal ordering.
In addition, we show that the same idea is also viable for thermal leptogenesis
in a {\it minimal} type-I seesaw model \cite{Frampton:2002qc,Guo:2006qa}
with two nearly degenerate heavy Majorana neutrinos.

\section{Resonant leptogenesis}

In the type-I seesaw scenario the lepton-number-violating decays
$N^{}_i \to \ell^{}_\alpha + H$ and $N^{}_i \to \overline{\ell^{}_\alpha} +
\overline{H}$ are also CP-violating, thanks to the interference between
their tree and one-loop (self-energy and vertex-correction) amplitudes \cite{Fukugita:1986hr,Luty:1992un,Covi:1996wh,Plumacher:1996kc}. Given
$M^{}_1 \simeq M^{}_2 \ll M^{}_3$, however, the near degeneracy of $M^{}_1$ and
$M^{}_2$ can make the one-loop self-energy contribution {\it resonantly}
enhanced \cite{Pilaftsis:1997dr,Pilaftsis:1997jf,Pilaftsis:2003gt,Anisimov:2005hr,Xing:2006ms,Dev:2014laa,Zhang:2015tea,Bambhaniya:2016rbb,Dev:2017wwc}.
As a result, the flavor-dependent CP-violating asymmetries $\varepsilon^{}_{i\alpha}$
between $N^{}_i \to \ell^{}_\alpha + H$ and $N^{}_i \to \overline{\ell^{}_\alpha} +
\overline{H}$ decays (for $i=1,2$ and $\alpha = e, \mu, \tau$)
are dominated by the interference effect associated with
the self-energy diagram \cite{Pilaftsis:1997jf,Pilaftsis:2003gt}:
\begin{eqnarray}
\varepsilon^{}_{i\alpha} \hspace{-0.2cm} & \equiv & \hspace{-0.2cm}
\frac{\displaystyle \Gamma \left( N^{}_{i} \to \ell^{}_{\alpha} + H \right) - \Gamma
\left( N^{}_{i} \to \overline{\ell^{}_{\alpha}} + \overline{H} \right)}
{ \displaystyle \sum_\alpha \left[ \Gamma \left( N^{}_{i} \to \ell^{}_{\alpha}
+ H \right) + \Gamma \left( N^{}_{i} \to \overline{\ell^{}_{\alpha}} +
\overline{H} \right) \right]}
\nonumber \\
\hspace{-0.2cm} & = & \hspace{-0.2cm}
\frac{ \displaystyle  {\rm Im} \left[ \left(Y^{*}_\nu\right)^{}_{\alpha i}
\left(Y^{}_\nu\right)^{}_{\alpha j} \left(Y^\dagger_\nu Y^{}_\nu\right)^{}_{ij}
+ \xi^{}_{ij} \left(Y^{*}_\nu\right)^{}_{\alpha i} \left(Y^{}_\nu\right)^{}_{\alpha j}
\left(Y^\dagger_\nu Y^{}_\nu\right)^{}_{ji} \right]}{\displaystyle\left(Y^\dagger_\nu
Y^{}_\nu\right)^{}_{ii} \left(Y^\dagger_\nu Y^{}_\nu \right)^{}_{jj}}
\cdot \frac{\xi^{}_{ij} \zeta^{}_j \left(\xi^2_{ij} -1\right)}
{\left(\xi^{}_{ij} \zeta^{}_j\right)^2 + \left(\xi^2_{ij} -1\right)^2} \; , \hspace{0.6cm}
%   (7)
\end{eqnarray}
where $\xi^{}_{ij} \equiv M^{}_i/M^{}_j$ and
$\zeta^{}_j \equiv \left(Y^\dagger_\nu Y^{}_\nu \right)^{}_{jj}/\left(8\pi\right)$
with the Latin subscripts $j\neq i$ running over $1$ and $2$.
Taking account of the expression of $Y^{}_\nu$ in Eq.~(6), we immediately arrive at
\begin{eqnarray}
\left(Y^{*}_\nu\right)^{}_{e i} \left(Y^{}_\nu\right)^{}_{e j}
\hspace{-0.2cm} & = & \hspace{-0.2cm} \frac{I^2_0}{v^2}
\sqrt{m^{}_i m^{}_j M^{}_i M^{}_j} \hspace{0.05cm} U^\ast_{e i} U^{}_{e j} \; ,
\nonumber \\
\left(Y^{*}_\nu\right)^{}_{\mu i} \left(Y^{}_\nu\right)^{}_{\mu j}
\hspace{-0.2cm} & = & \hspace{-0.2cm} \frac{I^2_0}{v^2} \sqrt{m^{}_i m^{}_j
M^{}_i M^{}_j} \hspace{0.05cm} U^\ast_{\mu i} U^{}_{\mu j} \; ,
\nonumber \\
\left(Y^{*}_\nu\right)^{}_{\tau i} \left(Y^{}_\nu\right)^{}_{\tau j}
\hspace{-0.2cm} & = & \hspace{-0.2cm} \frac{I^2_0}{v^2} \left(1 + 2\Delta^{}_\tau\right)
\sqrt{m^{}_i m^{}_j M^{}_i M^{}_j} \hspace{0.05cm} U^\ast_{\tau i} U^{}_{\tau j}
+ {\cal O} \left( \Delta^2_\tau \right) \; ,
%     (8)
\end{eqnarray}
together with
\begin{eqnarray}
\left(Y^\dagger_\nu Y^{}_\nu\right)^{}_{ij} \hspace{-0.2cm} & = & \hspace{-0.2cm}
\frac{I^2_0}{v^2} \sqrt{m^{}_i m^{}_j M^{}_i M^{}_j} \left( \delta^{}_{ij} +
2\Delta^{}_\tau U^\ast_{\tau i} U^{}_{\tau j} \right) + \mathcal{O}
\left( \Delta^2_\tau \right) \; .
%   (9)
\end{eqnarray}
The flavored CP-violating asymmetries in Eq.~(7) turn out to be
\begin{eqnarray}
\varepsilon^{}_{i\alpha} = 2\Delta^{}_\tau \left[ {\rm Im}
\left( U^\ast_{\tau i} U^{}_{\tau j} U^\ast_{\alpha i} U^{}_{\alpha j}
\right) + \xi^{}_{ij} {\rm Im} \left( U^\ast_{\tau j}
U^{}_{\tau i} U^\ast_{\alpha i} U^{}_{\alpha j} \right) \right]
\frac{\xi^{}_{ij} \zeta^{}_j \left(\xi^2_{ij} -1\right)}
{\left(\xi^{}_{ij} \zeta^{}_j\right)^2 + \left(\xi^2_{ij} -1\right)^2}
+ \mathcal{O} \left( \Delta^2_\tau \right) \; ,
%   (10)
\end{eqnarray}
where $\alpha = e, \mu, \tau$ and $j\neq i =1,2$; and $\zeta^{}_j
= I^2_0 \left(1+2\Delta^{}_\tau |U^{}_{\tau j}|^2\right)
m^{}_j M^{}_j/ \left(8\pi v^2\right) + \mathcal{O} \left( \Delta^2_\tau \right)$.
One can see that $\varepsilon^{}_{i\alpha} \propto \Delta^{}_\tau$ holds,
and hence $\varepsilon^{}_{i\alpha}$ will be vanishing or vanishingly small if
$O = {\bf 1}$ is taken but the RGE-induced effect is neglected.
Note that the first term in the square brackets of Eq.~(10) depends only
on a single combination of the two so-called Majorana phases $\rho$ and
$\sigma$ of $U$ \cite{Xing:2019vks}, denoted here as $\phi \equiv \rho-\sigma$;
and the second term is only dependent on the Dirac phase $\delta$ of $U$.
So a direct connection between the effects of leptonic CP violation at
high- and low-energy scales has been established in our RGE-assisted
seesaw-plus-leptogenesis scenario.

In the flavored resonant thermal leptogenesis scenario under consideration,
the CP-violating asymmetries $\varepsilon^{}_{i\alpha}$ are linked to the
baryon-to-photon ratio $\eta$ as follows \cite{Buchmuller:2002rq,Buchmuller:2003gz}:
\begin{eqnarray}
\eta \simeq -9.6\times 10^{-3} \sum_\alpha \left(\varepsilon^{}_{1\alpha}
\kappa^{}_{1\alpha} + \varepsilon^{}_{2\alpha} \kappa^{}_{2\alpha}\right) \;,
%   (11)
\end{eqnarray}
where $\kappa^{}_{1\alpha}$ and $\kappa^{}_{2\alpha}$ are the conversion efficiency
factors, and the sum over the flavor index $\alpha$ depends on which region
the lepton flavor(s) can take effect. To evaluate the sizes of $\kappa^{}_{i\alpha}$,
let us first of all figure out the effective light neutrino masses
\begin{eqnarray}
\widetilde{m}^{}_{i\alpha} \equiv \frac{ v^2 \left| \left( Y^{}_\nu
\right)^{}_{\alpha i} \right|^2 }{M^{}_i} = I^2_0 \left( 1+ 2\Delta^{}_\tau
\delta^{}_{\alpha\tau} \right) m^{}_i |U^{}_{\alpha i}|^2 +
{\cal O}\left(\Delta^2_\tau\right) \;.
%   (12)
\end{eqnarray}
Then the so-called decay parameters
$K^{}_{i\alpha} \equiv \widetilde{m}^{}_{i\alpha}/m^{}_\ast$ can be defined
and calculated,
where $m^{}_\ast = 8\pi v^2 H(M^{}_1) /M^2_1 \simeq 1.08\times 10^{-3}$ eV
represents the equilibrium neutrino mass and
$H(M^{}_1) = \sqrt{8\pi^3 g^{}_\ast/90} M^2_1/M^{}_{\rm pl}$ is the Hubble
expansion parameter of the Universe at temperature $T \simeq M^{}_1$ with
$g^{}_\ast =106.75$ being the total number of relativistic degrees of freedom
in the SM and $M^{}_{\rm pl} = 1.22\times10^{19}$ GeV being the Planck mass.
\begin{itemize}
\item     For $M^{}_i \gtrsim 10^{12}$ GeV (for $i =1,2$),
all the leptonic Yukawa interactions are flavor-blind.
In this case the {\it unflavored} leptogenesis depends on the overall
CP-violating asymmetry $\varepsilon^{}_i = \varepsilon^{}_{i e} +
\varepsilon^{}_{i \mu} + \varepsilon^{}_{i \tau} \simeq 0$ in our scenario,
as one can easily see from Eq.~(10).

\item     For $10^9~ {\rm GeV}~ \lesssim M^{}_i \lesssim 10^{12}$ GeV, the
$\tau$-flavored Yukawa interaction is in thermal equilibrium and thus the
$\tau$ flavor can be distinguished from $e$ and $\mu$ flavors in the
Boltzmann equations \cite{Buchmuller:2002rq,Buchmuller:2003gz}.
In this case one has to consider two classes of lepton flavors:
the $\tau$ flavor and a combination of the indistinguishable $e$ and $\mu$
flavors. We are then left with the flavored CP-violating asymmetries
$\varepsilon^{}_{i\tau}$ and $\varepsilon^{}_{ie} + \varepsilon^{}_{i\mu}$
together with the flavored decay parameters $K^{}_{i\tau}$ and
$K^{}_{ie} + K^{}_{i\mu}$, and the latter can be used to determine the
corresponding conversion efficiency factors.

\item     For $10^5~ {\rm GeV}~ \lesssim M^{}_i \lesssim 10^{9}$ GeV, the
$\mu$- and $\tau$-flavored Yukawa interactions are both in thermal equilibrium,
making the $\mu$ and $\tau$ flavors distinguishable. That is why
all the three lepton flavors should be separately treated in this case.
\end{itemize}
Now that we are dealing with resonant leptogenesis, let us define a
dimensionless parameter $d \equiv \left( M^{}_2 - M^{}_1\right)/M^{}_1
= \xi^{}_{21} -1$ to measure the level of degeneracy for two of the
three heavy Majorana neutrinos. Allowing for $d \ll 1$, we have
$\kappa^{}_{1\alpha} \simeq \kappa^{}_{2\alpha} \equiv \kappa
\left( K^{}_\alpha \right)$ with
$K^{}_\alpha \equiv K^{}_{1\alpha} + K^{}_{2\alpha}$\cite{Blanchet:2006be,Blanchet:2006dq}.
Given the initial thermal abundance of heavy Majorana neutrinos, the
efficiency factor $\kappa \left( K^{}_\alpha \right)$ can be approximately
expressed as \cite{Buchmuller:2004nz,Blanchet:2006be}
\begin{eqnarray}
\kappa \left( K^{}_\alpha \right) \simeq \frac{2}{K^{}_\alpha z^{}_{\rm B}
\left( K^{}_\alpha \right)} \left[ 1 - \exp \left( -\frac{1}{2} K^{}_\alpha z^{}_{\rm B}
\left( K^{}_\alpha \right) \right) \right] \;,
%   (13)
\end{eqnarray}
where $z^{}_{\rm B} \left( K^{}_\alpha \right) \simeq 2 + 4K^{0.13}_\alpha \exp
\left( -2.5/K^{}_\alpha \right)$.
%%%%%%%%%%%%%%%%%%%%%%%%%%%%%%%% figure 1 %%%%%%%%%%%%%%%%%%%%%%%%%%%%%%%%%
\begin{figure}[t]
  \centering
  \includegraphics[width=16.5cm]{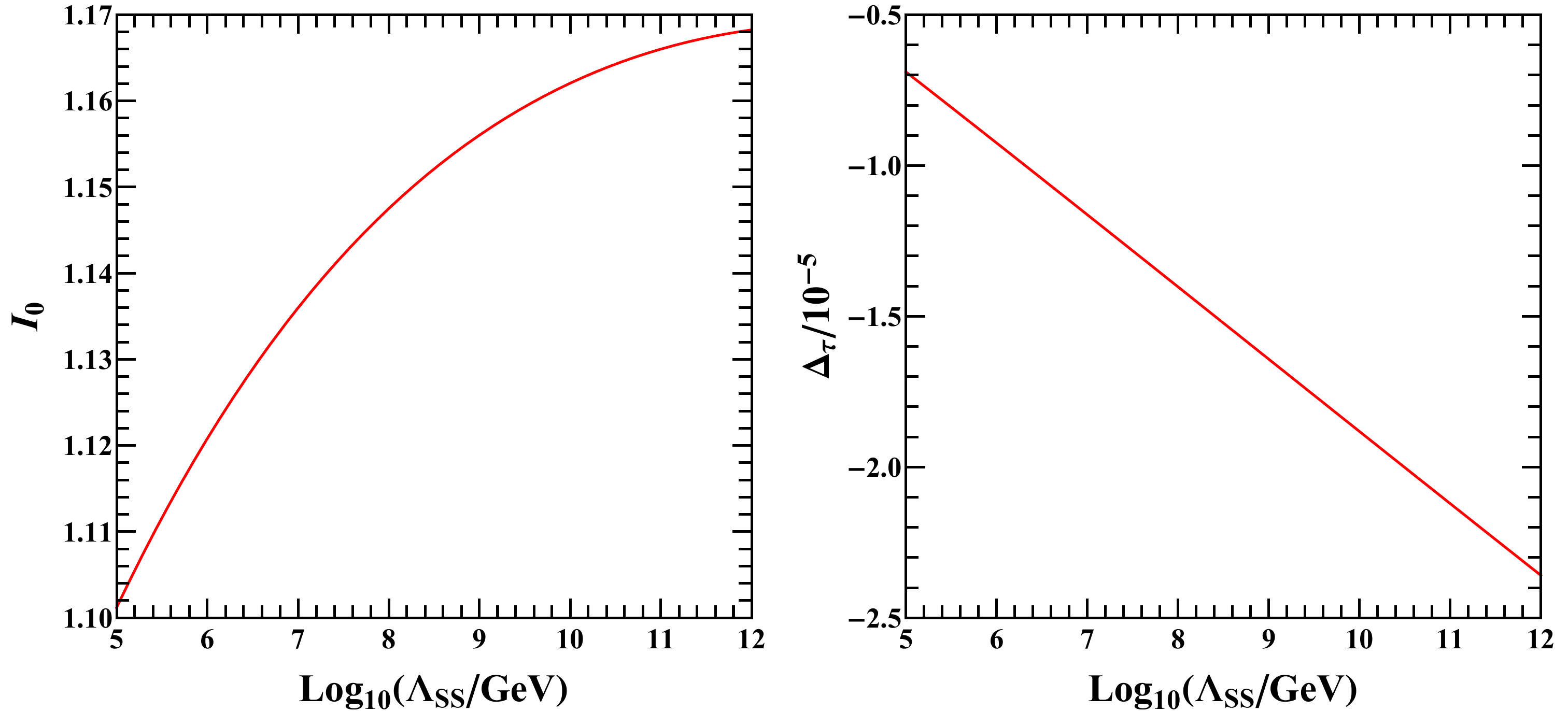}\\
  \caption{The values of $I^{}_0$ and $\Delta^{}_\tau$ against the seesaw scale
  $\Lambda^{}_{\rm SS} \in [10^{5}, 10^{12}]$ GeV in the SM.}\label{I-Delta}
\end{figure}
%%%%%%%%%%%%%%%%%%%%%%%%%%%%%%%%%%%%%%%%%%%%%%%%%%%%%%%%%%%%%%%%%%%%%%%%%%%

We proceed to numerically illustrate that our resonant leptogenesis scenario works
well. First of all, the values of $I^{}_0$ and $\Delta_\tau$ at the seesaw scale
are illustrated in Fig.~\ref{I-Delta} with $\Lambda^{}_{\rm SS} \in [10^{5}, 10^{12}]$
GeV in the SM. Adopting the standard parametrization of $U$ \cite{Xing:2019vks},
we need to input the values of eleven parameters: two heavy neutrino masses
$M^{}_1$ and $M^{}_2$ (or equivalently, $M^{}_1$ and $d$); three light
neutrino masses $m^{}_i$ (for $i=1,2,3$); three lepton flavor mixing angles
$\theta^{}_{12}$, $\theta^{}_{13}$ and $\theta^{}_{23}$; and three CP-violating phases
$\delta$, $\rho$ and $\sigma$ (but only $\delta$ and the combination
$\phi \equiv \rho-\sigma$ contribute). For the sake of simplicity, here
we only input the best-fit values of $\theta^{}_{12}$, $\theta^{}_{13}$,
$\theta^{}_{23}$, $\delta$, $\Delta m^2_{21} \equiv m^2_2 - m^2_1$ and
$\Delta m^2_{31} \equiv m^2_3 - m^2_1$ (or $\Delta m^2_{32} \equiv m^2_3 - m^2_2$)
extracted from a recent global analysis of current neutrino oscillation
data \cite{Capozzi:2018ubv,Esteban:2018azc}:
$\sin^2 \theta^{}_{12} = 0.310$, $\sin^2 \theta^{}_{13} = 0.02241$ (or $0.02261$),
$\sin^2 \theta^{}_{23} = 0.558$ (or $0.563$), $\delta = 222^\circ$ (or $285^\circ$),
$\Delta m^2_{21} = 7.39\times 10^{-5} ~{\rm eV^2}$ and $\Delta m^2_{31} =
2.523 \times 10^{-3} ~{\rm eV^2}$ (or $\Delta m^2_{32} = -2.509 \times 10^{-3}
~{\rm eV^2}$) for the normal (or inverted) neutrino mass ordering. Then we are left
with only four unknown parameters: $m^{}_1$ (or $m^{}_3$), $M^{}_1$, $d$ and
$\phi$.

Given the above inputs, we can estimate the size of $K^{}_{\alpha}$ with the help
of Eq.~(12). It is found that $K^{}_e \gtrsim 2.4$, $K^{}_\mu \gtrsim 2.9$
and $K^{}_\tau \gtrsim 2.6$ in the normal neutrino mass ordering case; or
$K^{}_e \gtrsim 44.9$, $K^{}_\mu \gtrsim 20.8$ and $K^{}_\tau \gtrsim 26.4$
in the inverted mass ordering case. Now that $K^{}_\alpha > 1$ holds in either
case, any lepton-antilepton asymmetries generated by the lepton-number-violating
and CP-violating decays of $N^{}_3$ with $M^{}_3 \gg M^{}_1 \simeq M^{}_2$
can be efficiently washed out. It is therefore safe to only consider the
asymmetries produced by the decays of $N^{}_1$ and $N^{}_2$.
%%%%%%%%%%%%%%%%%%%%%%%%%%%%%%%% figure 2 %%%%%%%%%%%%%%%%%%%%%%%%%%%%%%%%%
\begin{figure}[h!]
  \centering
  \includegraphics[width=16.5cm]{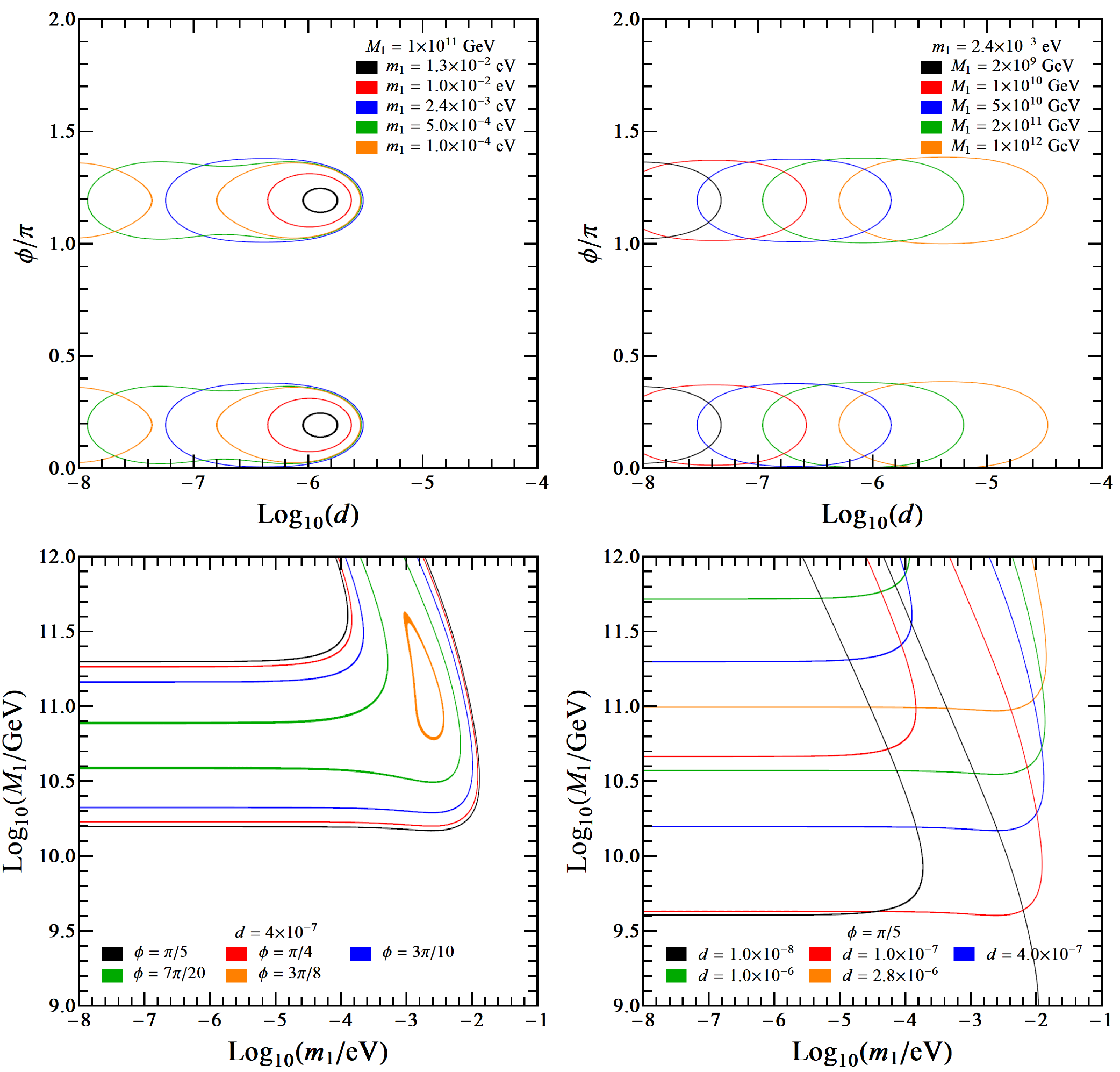} \\
  \caption{A viable RGE-assisted resonant leptogenesis scenario for the $\tau$-flavored
  regime with temperature $T \simeq M^{}_1 \in \left( 10^9, 10^{12} \right]$
  GeV in the {\it normal} neutrino mass ordering case: the parameter space of
  $d$ and $\phi$ (upper panels) with some given values of $m^{}_1$ and
  $M^{}_1$; and the parameter space of $m^{}_1$ and $M^{}_1$ (lower panels)
  with some given values of $d$ and $\phi$.}\label{two-flavor}
\end{figure}
%%%%%%%%%%%%%%%%%%%%%%%%%%%%%%%%%%%%%%%%%%%%%%%%%%%%%%%%%%%%%%%%%%%%%%%%%%%

\begin{figure}[h!]
  \centering
  \includegraphics[width=16.5cm]{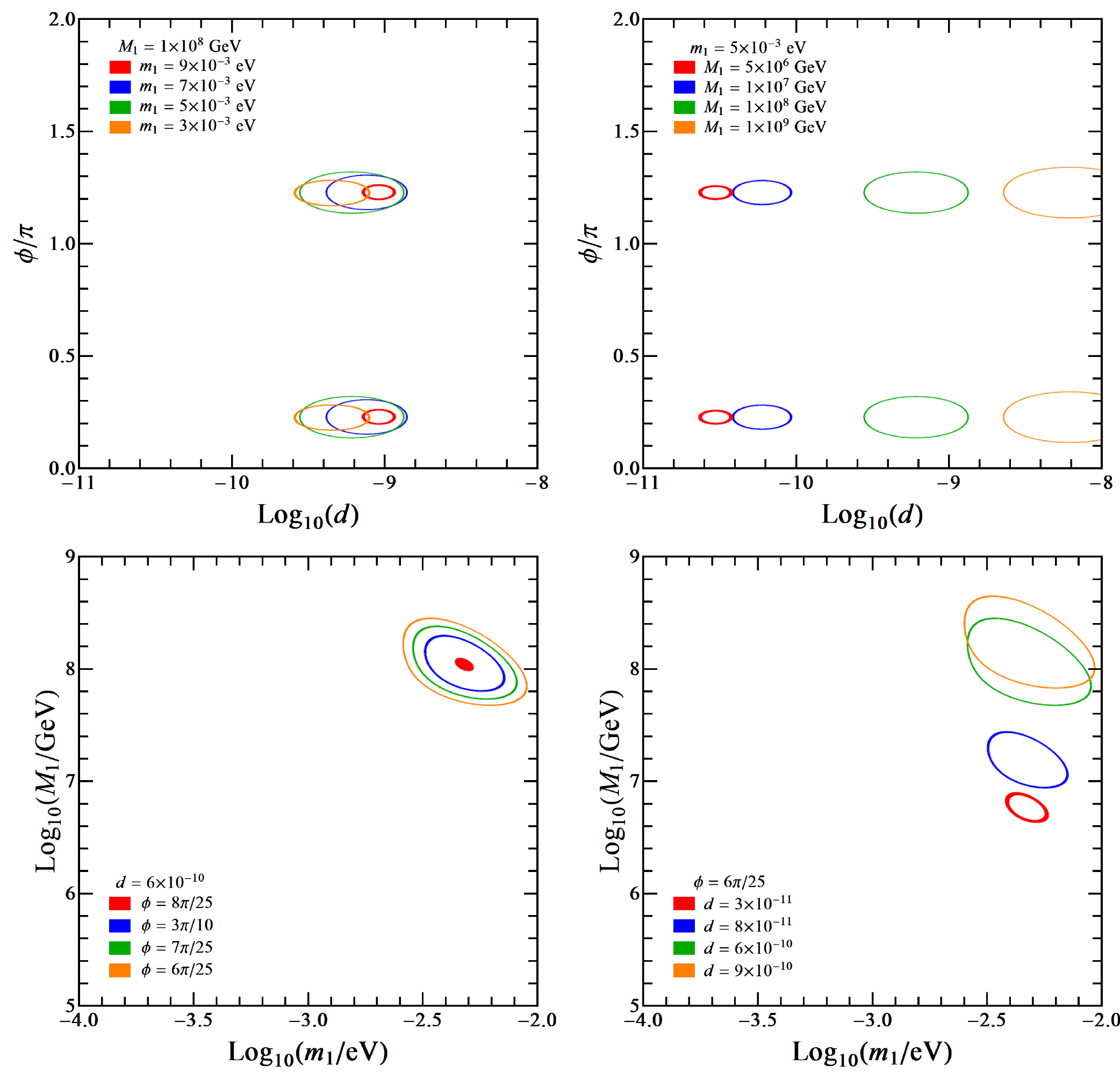}\\
  \caption{A viable RGE-assisted resonant leptogenesis scenario for the
  $(\mu+\tau)$-flavored regime with temperature
  $T \simeq M^{}_1 \in \left( 10^5, 10^9 \right]$
  GeV in the {\it normal} neutrino mass ordering case: the parameter space of
  $d$ and $\phi$ (upper panels) with some given values of $m^{}_1$ and
  $M^{}_1$; and the parameter space of $m^{}_1$ and $M^{}_1$ (lower panels)
  with some given values of $d$ and $\phi$.}\label{three-flavor}
\end{figure}

Now let us use the observed value of $\eta$ to constrain the parameter
space of $\phi$ and $d$ by allowing $m^{}_1$ (or $m^{}_3$) and $M^{}_1$
to vary in some specific ranges; or to constrain the parameter space
of $m^{}_1$ (or $m^{}_3$) and $M^{}_1$ by allowing $\phi$ and $d$
to vary in some specific ranges, and by taking account of both the
$\tau$-flavored regime with $T \simeq M^{}_1 \in \left(10^9, 10^{12}
\right]$ GeV and the $(\mu+\tau)$-flavored regime with $T \simeq M^{}_1
\in \left( 10^5, 10^9 \right]$ GeV. We find no parameter space in the
inverted neutrino mass ordering case, in which the conversion efficiency
factors are strongly suppressed. Our RGE-assisted resonant leptogenesis
scenario is viable in the normal neutrino mass ordering case,
and the numerical results for the $\tau$- and
$(\mu+\tau)$-flavored regimes are shown in Figs.~\ref{two-flavor} and
\ref{three-flavor}, respectively. Some brief discussions are in order.
%%%%%%%%%%%%%%%%%%%%%%%%%%%%%%%% figure 3 %%%%%%%%%%%%%%%%%%%%%%%%%%%%%%%%%
%%%%%%%%%%%%%%%%%%%%%%%%%%%%%%%%%%%%%%%%%%%%%%%%%%%%%%%%%%%%%%%%%%%%%%%%%%%
\begin{itemize}
\item       {\it The $\tau$-flavored regime} (i.e., $T \simeq M^{}_1
\in \left( 10^9, 10^{12} \right]$ GeV). As can be seen in the upper panels
of Fig.~\ref{two-flavor}, $\phi$ is mainly allowed to lie in two possible
ranges: $[0,2\pi/5]$ and $[\pi, 7\pi/5]$; and the dimensionless parameter
$d$ satisfies $d \lesssim 4\times 10^{-5}$. These two ranges of
$\phi$ differ from each other just by a shift or reflection; and they
are symmetric about $\phi = \pi/5$ and $\phi = 6\pi/5$, respectively.
Such a feature can easily be understood. Considering
$\varepsilon^{}_{i e} + \varepsilon^{}_{i \mu} + \varepsilon^{}_{i \tau} =0$
and $M^{}_1 \simeq M^{}_2$, we have $\eta \propto \varepsilon^{}_{1\tau}
+ \varepsilon^{}_{2\tau} \propto \sin{2(\phi - \varphi^{}_\tau)}$ with
$\varphi^{}_\tau \equiv \arg{\left(U^\ast_{\tau 1} U^{}_{\tau2}
e^{{\rm i}\phi}\right)}$ being dominated by the CP-violating phase $\delta$
whose value is around $19\pi/20$. And thus if $\phi$ is replaced by
$\pi + \phi$ (or $7\pi/5 - \phi$) and $2\pi/5 - \phi$ (or $12\pi/5 - \phi$),
the value of $\eta$ will keep unchanged. Note that even if $\phi = 0$ holds,
there can still exist some parameter space for the four free parameters.
In this special case the Dirac CP phase $\delta$, which is sensitive to
leptonic CP violation in neutrino oscillations, is the only source of CP violation
in our flavored resonant leptogenesis scenario. As shown in the lower panels of
Fig.~\ref{two-flavor}, $M^{}_1$ varies in the range
$\left( 10^9, 10^{12} \right]$ GeV and $m^{}_1 \lesssim 0.01$ eV holds.
But for a given value of $d$, the parameter space of $M^{}_1$ is
generally constrained to a specific range; and when $d$ decreases, the
allowed range of $M^{}_1$ increases correspondingly. For the most part of the
allowed range of $\phi$, the smallest neutrino mass $m^{}_1$ can approach
zero with a given value of $d$, and the value of $\eta$ is almost
independent of $m^{}_1$ when $m^{}_1$ becomes small enough since $\eta$
is dominated by the term containing $m^{}_2 \gg m^{}_1$ in this case.
When the value of $\phi$ approaches the edge of the allowed range of $\phi$,
there will be a lower limit on $m^{}_1$ which can be seen from the orange
band in the lower-left panel of Fig.~\ref{two-flavor}. This feature is mainly
a consequence of the reduction of $\varepsilon^{}_{i\tau}$ in magnitude,
which is proportional to $\sin{2(\phi - \varphi^{}_\tau)}$.

\item       {\it The $(\mu+\tau)$-flavored regime} (i.e., $T \simeq
M^{}_1 \in \left( 10^5, 10^9 \right]$ GeV). It is obvious that in this case
the parameter space is largely reduced as compared with that in the
$\tau$-flavored regime. The main reason is that there exists a large
cancellation among the contributions of three flavors; namely, the terms
$\left(\varepsilon^{}_{1\alpha} + \varepsilon^{}_{2\alpha}\right)
\kappa (K^{}_\alpha)$ (for $\alpha=e,\mu,\tau$) may cancel one another.
As shown in the upper panels of Fig.~\ref{three-flavor}, $\phi$ is
mainly located in the intervals $\left[ \pi/10, 8\pi/25 \right]$ and
$\left[ 11\pi/10, 33\pi/25 \right]$, but it cannot vanish. The value
of $d$ is strongly suppressed, and it mainly lies in the range
$\left[2\times10^{-11}, 10^{-8} \right]$. The two intervals of $\phi$ have
quite similar properties as those in the $\tau$-flavored regime, but
their symmetry axes are determined by the interference between
$\varepsilon^{}_{i\mu} \propto \sin{(2\phi - \varphi^{}_\mu-\varphi^{}_\tau)}$
and $\varepsilon^{}_{i\tau} \propto \sin{2(\phi - \varphi^{}_\tau)}$ with
$\varphi^{}_\alpha \equiv \arg{\left(U^\ast_{\alpha 1} U^{}_{\alpha 2}
e^{{\rm i}\phi}\right)}$ (for $\alpha = \mu, \tau$). The parameter space
of $m^{}_1$ and $M^{}_1$ is mainly described by
$m^{}_1 \in \left[2.5\times10^{-3},10^{-2} \right]$ eV
and $M^{}_1 \in \left[4\times 10^6, 10^9\right]$ GeV.
\end{itemize}
It is finally worth mentioning that the normal neutrino mass ordering is
currently favored over the inverted one at the $3\sigma$ level, as indicated by
a global analysis of today's available experimental data on various neutrino
oscillation phenomena \cite{Capozzi:2018ubv,Esteban:2018azc,Vagnozzi:2017ovm}.
This indication is certainly consistent with our RGE-assisted resonant
leptogenesis scenario.

\section{On the minimal seesaw}

\begin{figure}[h!]
  \centering
  \includegraphics[width=16.5cm]{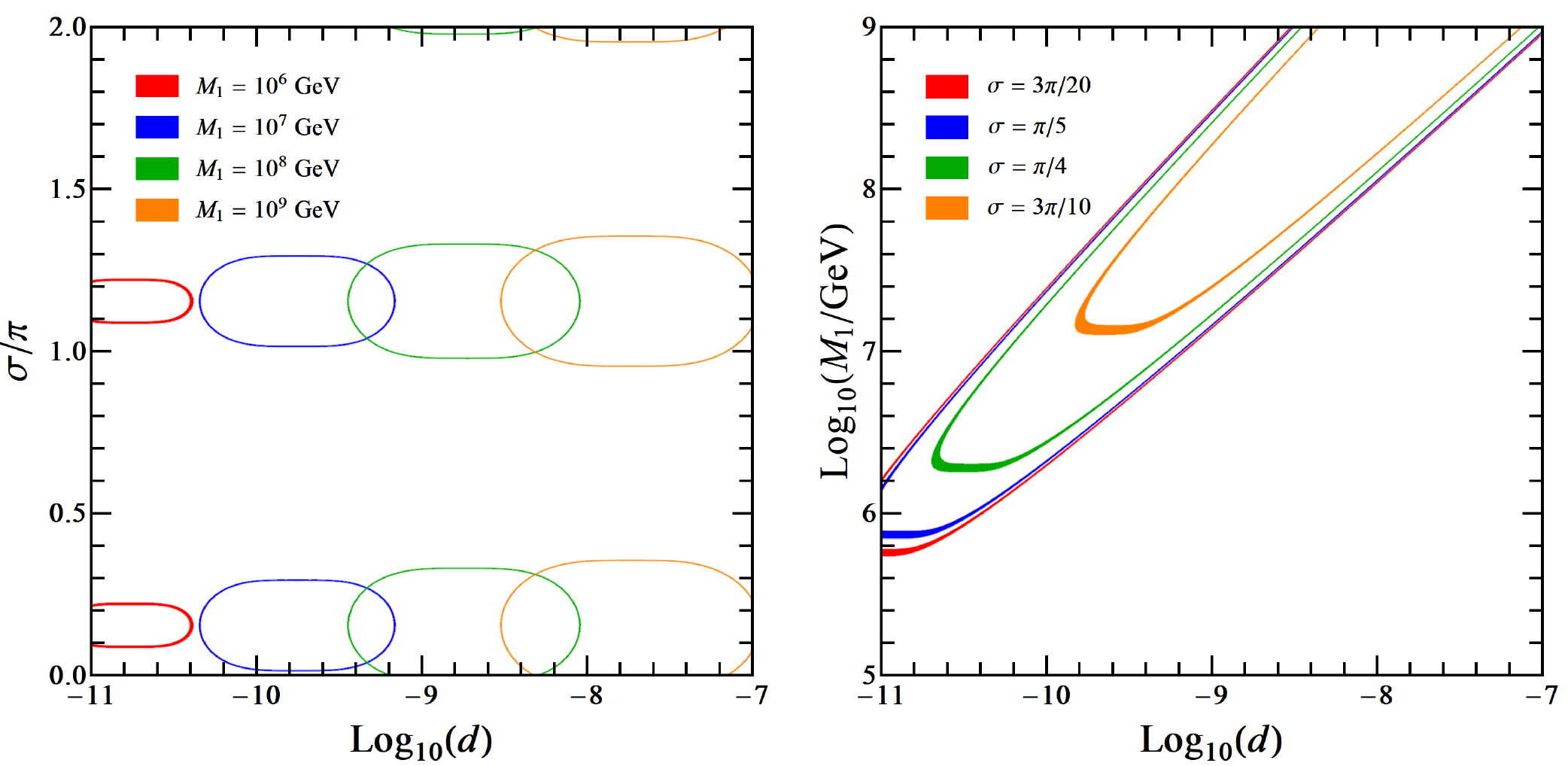}\\
  \caption{A viable RGE-assisted resonant leptogenesis scenario for the
  $(\mu+\tau)$-flavored regime with temperature $T \simeq M^{}_1
  \in \left( 10^5, 10^9 \right]$ GeV in the {\it normal} neutrino mass
  ordering case based on the minimal type-I seesaw: the parameter
  space of $d$ and $\sigma$ (left panel) with some given values
  of $M^{}_1$; and the parameter space of $d$ and $M^{}_1$
  (right panel) with some given values of $\sigma$.}\label{Min}
\end{figure}

Since we have focused on resonant leptogenesis with $M^{}_1 \simeq M^{}_2
\ll M^{}_3$ based on the type-I seesaw mechanism, it is natural to consider
a minimized version of this scenario by switching off the heaviest Majorana
neutrino $N^{}_3$. That is, we can simply invoke the minimal type-I seesaw
model \cite{Frampton:2002qc,Guo:2006qa} with two nearly degenerate heavy
Majorana neutrinos to realize resonant leptogenesis. In this case the Yukawa
coupling matrix is a $3\times 2$ matrix, and thus the arbitrary orthogonal
matrix $O$ in the CI parametrization of $Y^{}_\nu$ is also a $3\times 2$
matrix. To remove the uncertainties associated with $O$, we may take
\begin{eqnarray}
O = \left(\begin{matrix} 0 & 0 \\ 1 & 0 \\ 0 & 1 \end{matrix}\right) \;,
\quad {\rm or} \quad O = \left(\begin{matrix} 1 & 0 \\ 0 & 1 \\ 0 & 0
\end{matrix}\right) \;,
%   (14)
\end{eqnarray}
corresponding to the normal ($m^{}_1 =0$) or inverted ($m^{}_3 =0$) neutrino
mass ordering. Then the expression of $Y^{}_\nu$ in Eq.~(6) can be simplified
to
\begin{eqnarray}
Y^{}_\nu \hspace{-0.2cm} & = & \hspace{-0.2cm}
\frac{\rm i}{v} \hspace{0.05cm} I^{}_0  \left[\left(\begin{matrix}
\sqrt{m^{}_2 M^{}_1} \hspace{0.05cm} U^{}_{e 2} &
\sqrt{m^{}_3 M^{}_2} \hspace{0.05cm} U^{}_{e 3} \cr
\sqrt{m^{}_2 M^{}_1} \hspace{0.05cm} U^{}_{\mu 2} &
\sqrt{m^{}_3 M^{}_2} \hspace{0.05cm} U^{}_{\mu 3} \cr
\sqrt{m^{}_2 M^{}_1} \hspace{0.05cm} U^{}_{\tau 2} &
\sqrt{m^{}_3 M^{}_2} \hspace{0.05cm} U^{}_{\tau 3} \cr
\end{matrix}\right) +
\Delta^{}_\tau \left(\begin{matrix}
0 & 0 \cr 0 & 0 \cr
\sqrt{m^{}_2 M^{}_1} \hspace{0.05cm} U^{}_{\tau 2} &
\sqrt{m^{}_3 M^{}_2} \hspace{0.05cm} U^{}_{\tau 3} \cr
\end{matrix}\right) \right] \;
%   (15)
\end{eqnarray}
with $m^{}_1 =0$, $m^{}_2 =\sqrt{\Delta m^2_{21}}$ and $m^{}_3 =\sqrt{\Delta m^2_{31}}$;
or
\begin{eqnarray}
Y^{}_\nu \hspace{-0.2cm} & = & \hspace{-0.2cm}
\frac{\rm i}{v} \hspace{0.05cm} I^{}_0 \left[\left(\begin{matrix}
\sqrt{m^{}_1 M^{}_1} \hspace{0.05cm} U^{}_{e 1} &
\sqrt{m^{}_2 M^{}_2} \hspace{0.05cm} U^{}_{e 2} \cr
\sqrt{m^{}_1 M^{}_1} \hspace{0.05cm} U^{}_{\mu 1} &
\sqrt{m^{}_2 M^{}_2} \hspace{0.05cm} U^{}_{\mu 2} \cr
\sqrt{m^{}_1 M^{}_1} \hspace{0.05cm} U^{}_{\tau 1} &
\sqrt{m^{}_2 M^{}_2} \hspace{0.05cm} U^{}_{\tau 2} \cr
\end{matrix}\right) +
\Delta^{}_\tau \left(\begin{matrix}
0 & 0 \cr 0 & 0 \cr
\sqrt{m^{}_1 M^{}_1} \hspace{0.05cm} U^{}_{\tau 1} &
\sqrt{m^{}_2 M^{}_2} \hspace{0.05cm} U^{}_{\tau 2} \cr
\end{matrix}\right) \right] \;
%   (16)
\end{eqnarray}
with $m^{}_3 =0$, $m^{}_2 =\sqrt{-\Delta m^2_{32}}$ and
$m^{}_1 =\sqrt{-\Delta m^2_{32}-\Delta m^2_{21}}$. In other words, the mass
spectrum of three light neutrinos is fully fixed by current neutrino oscillation
data in the minimal seesaw model, so the uncertainty associated with the
absolute light neutrino mass scale disappears. Another bonus is that
one of the Majorana phases of $U$ (i.e., $\rho$) can always be removed thanks
to the vanishing of $m^{}_1$ or $m^{}_3$, and therefore we are left with only
two low-energy CP-violating phases (i.e., $\delta$ and $\sigma$) which affect
the flavored CP-violating asymmetries $\varepsilon^{}_{i\alpha}$.
In our numerical calculations we simply input the best-fit values of
$\theta^{}_{12}$, $\theta^{}_{13}$, $\theta^{}_{23}$, $\delta$,
$\Delta m^2_{21}$ and $\Delta m^2_{31}$ (or $\Delta m^2_{32}$) as given
below Eq.~(13). Then the observed value of $\eta$ can be used to constrain
the parameter space of $\sigma$ and $d$ by allowing $M^{}_1$ to
vary in some specific ranges; or to constrain the parameter
space of $M^{}_1$ and $d$ by allowing $\sigma$ to vary in $\left(0,2\pi
\right]$. We find that in this minimal type-I seesaw model our RGE-assisted resonant
leptogenesis scenario is viable only for the normal neutrino mass ordering
with $m^{}_1 =0$ and only in the $(\mu+\tau)$-flavored regime. The numerical
results are briefly illustrated in Fig.~\ref{Min}.
%%%%%%%%%%%%%%%%%%%%%%%%%%%%%%%% figure 4 %%%%%%%%%%%%%%%%%%%%%%%%%%%%%%%%%
%%%%%%%%%%%%%%%%%%%%%%%%%%%%%%%%%%%%%%%%%%%%%%%%%%%%%%%%%%%%%%%%%%%%%%%%%%%

An immediate comparison between Fig.~\ref{three-flavor} and Fig.~\ref{Min},
which are both associated with the $(\mu+\tau)$-flavored regime for
resonant leptogenesis, tells us that the parameter space in the minimal
seesaw case is slightly larger. This observation is attributed to the smaller
cancellation among the contributions of three flavors, since the
efficiency factor for the $e$ flavor [i.e., $\kappa(K^{}_e)$] is much larger
than those for $\mu$ and $\tau$ flavors [i.e., $\kappa(K^{}_\mu)$ and
$\kappa(K^{}_\tau)$] in the minimal seesaw scenario.
Note that if $\kappa(K^{}_e) = \kappa(K^{}_\mu) =
\kappa(K^{}_\tau)$ held, $\eta$ would vanish due to
$\varepsilon^{}_{i e} + \varepsilon^{}_{i \mu} + \varepsilon^{}_{i \tau} = 0$.
As shown in Fig.~\ref{Min}, $\sigma$ is mainly located in two
disconnected intervals $[0,3\pi/10]$ and $[\pi,13\pi/10]$.
But these two intervals are different from each other only by a shift
($\sigma \to \sigma + \pi$) or a reflection (about $\sigma = 13\pi/20$); and
each of them has a symmetry axis ($\sigma = 3\pi/20$ or $\sigma = 23\pi/20$).
We see that $M^{}_1 \gtrsim 6.3 \times 10^5$ GeV holds, and $d$ is allowed
to vary in a wide range between $10^{-11}$ and $10^{-7}$.
When the value of $M^{}_1$ deceases, the lower and upper bounds of $d$ are both
reduced; meanwhile, the allowed range of $\sigma$ becomes smaller.
That is why when $M^{}_1$ is smaller than $10^7$ GeV and $\sigma$ is switched off
(i.e., $\delta$ is the only source of CP violation),
it will be very difficult (and even impossible) to make our RGE-assisted
resonant leptogenesis scenario viable.

One may certainly extend the above ideas and discussions from the SM to the
MSSM, in which the magnitude of $\Delta^{}_\tau$ is expected to be enhanced
by taking a large value of $\tan \beta$. In this case it should be easier
to obtain more appreciable CP-violating asymmetries $\varepsilon^{}_{i\alpha}$,
simply because they are proportional to $\Delta^{}_\tau$. So a successful
RGE-assisted resonant leptogenesis can similarly be achieved in the MSSM case.
In this connection the main concern is how to avoid
the gravitino-overproduction problem
\cite{Khlopov:1984pf,Ellis:1984eq,Ellis:1984er,Asaka:2000zh,Davidson:2008bu},
and a simple way out might just be to require $M^{}_1 \lesssim 10^9$ GeV and focus
on thermal leptogenesis in the $(\mu+\tau)$-flavored regime.

\section{Summary}

Based on the type-I seesaw mechanism, we have reconstructed the Yukawa coupling
matrix $Y^{}_\nu$ in terms of the light Majorana neutrino masses $m^{}_i$, the
heavy Majorana neutrino masses $M^{}_i$ and the PMNS matrix $U$ by assuming the
arbitrary orthogonal matrix $O$ in the CI parametrization of $Y^{}_\nu$ to be
the identity matrix. To bridge the gap between $m^{}_i$ and $U$ at the seesaw
scale $\Lambda^{}_{\rm SS}$ and their counterparts at the Fermi scale
$\Lambda^{}_{\rm EW}$, we have taken into account the RGE-induced correction
to the light Majorana neutrino mass matrix. This RGE-modified seesaw formula
allows us to establish a direct link between low-energy CP violation and
flavored resonant leptogenesis with $M^{}_1 \simeq M^{}_2 \ll M^{}_3$, so as to
successfully interpret the observed baryon-antibaryon asymmetry of the Universe.
We have shown that our idea {\it does} work in either the $\tau$-flavored regime
with equilibrium temperature $T \simeq M^{}_1 \in (10^9, 10^{12}]$ GeV or
the $(\mu+\tau)$-flavored regime with $T \simeq M^{}_1 \in (10^5, 10^9]$ GeV,
provided the mass spectrum of three light Majorana neutrinos is normal rather
than inverted. We have also shown that the same idea is viable for a {\it minimal}
type-I seesaw model with two nearly degenerate heavy Majorana neutrinos.

\section*{Acknowledgements}

We would like to thank Zhen-hua Zhao and Shun Zhou for very useful discussions.
This work is supported in part by the National Natural
Science Foundation of China under grant No. 11775231 and grant No. 11835013.

%\newpage

\end{document}